\documentclass[a4paper, 10pt]{cas-sc}
\usepackage[authoryear]{natbib}
\def\tsc#1{\csdef{#1}{\textsc{\lowercase{#1}}\xspace}}
\tsc{WGM}
\tsc{QE}
\tsc{EP}
\tsc{PMS}
\tsc{BEC}
\tsc{DE}

\ExplSyntaxOn
\keys_set:nn { stm / mktitle } { nologo }
\ExplSyntaxOff

\newcommand{\significancestatement}[1]{}  %
\newcommand{\correspondingauthor}[1]{}
\newcommand{\authorcontributions}[1]{}
\newcommand{\authordeclaration}[1]{}
\newcommand{\equalauthors}[1]{}

\usepackage[australian]{babel}

\usepackage[utf8]{inputenc}
\usepackage{array}
\usepackage{mathtools}
\usepackage{amssymb}
\usepackage{amsmath}
\usepackage{amsfonts}
\usepackage{amsthm}
\usepackage{subfigure}
\usepackage{enumitem}
\usepackage{afterpage}
\usepackage{placeins}
\usepackage{siunitx}

\theoremstyle{plain}

\theoremstyle{definition}

\def\mrm{\langle a \rangle}
\DeclareMathOperator*{\negbin}{Neg
Bin}
\DeclareMathOperator*{\Acc}{Acc}

\renewcommand{\cite}[1]{\citep{#1}}

\begin{document}

\author[a]{{Ł}ukasz Brzozowski}[orcid=0000-0002-3625-3312]
\ead{lukasz.brzozowski@pw.edu.pl}
\cortext[cor1]{Corresponding author}
\cormark[1]
\credit{Conceptualisation of this study, Methodology, Investigation, Formal Analysis, Visualisation, Software, Writing}

\author[a,b]{Marek Gagolewski}[orcid=0000-0003-0637-6028]
\ead{marek.gagolewski@pw.edu.pl}
\ead[url]{https://www.gagolewski.com}
\credit{Conceptualisation of this study, Methodology, Investigation, Writing, Supervision}

\author[c]{Grzegorz Siudem}[orcid=0000-0002-9391-6477]
\ead{grzegorz.siudem@pw.edu.pl}
\ead[URL]{http://if.pw.edu.pl/~siudem}
\credit{Conceptualisation of this study, Methodology, Investigation, Formal Analysis, Writing}

\author[b]{Barbara Żogała-Siudem}[orcid=0000-0002-2869-7300]
\ead{zogala@ibspan.waw.pl}
\credit{Data Curation, Investigation, Software}

\shortauthors{Brzozowski, Gagolewski, Siudem, Żogała-Siudem}

\address[a]{Warsaw~University of Technology, Faculty of Mathematics and Information Science, ul. Koszykowa 75, 00-662 Warsaw, Poland}

\address[b]{Systems Research Institute, Polish Academy of Sciences,
ul. Newelska 6, 01-447 Warsaw, Poland}

\address[c]{Warsaw~University of Technology, Faculty of Physics,
ul. Koszykowa 75, 00-662 Warsaw, Poland}

\let\WriteBookmarks\relax
\def\floatpagepagefraction{1}
\def\textpagefraction{.001}

\shorttitle{The Price–Pareto growth model of networks with community structure}
\title[mode = title]{The Price–Pareto growth model of networks with community structure}

\begin{abstract}
We introduce a new analytical framework for modelling degree sequences in individual communities of real-world networks, e.g., citations to papers in different fields. Our work is inspired by a recent modification of the Price's model, which assumes that citations are gained partly accidentally, and to some extent preferentially. Our work addresses the need to represent the heterogeneity of various scientific domains, as standard homogeneous models fail to capture the distinct growth ratios and citing cultures of different fields. Extending the model to networks with a community structure allows us to devise the analytical formulae for, amongst others, citation counts in each cluster and their inequality as described by the Gini index. We also show that a citation count distribution in each community tends to a Pareto type II distribution. Thanks to the derived model parameter estimators, the new model can be fitted to real citation and similar networks.
\end{abstract}


\maketitle

\noindent{\small
Please cite this paper as:
Brzozowski, Ł., Gagolewski, M., Siudem, G., Żogała-Siudem, B., The Price–Pareto growth model of networks with community structure, \textit{Journal of Informetrics} \textbf{20}(3), 101850, 2026, \href{https://doi.org/DOI:10.1016/j.joi.2026.101850}{DOI:10.1016/j.joi.2026.101850}.
}

\section{Introduction}\label{sec:intro}
In recent years, there has been a surge of interest in analysing real-life network data, stimulated by the development of neural architectures for graph processing, such as Graph Neural Networks~\cite{scarselli2009} and Graph Convolutional Networks~\cite{kipfgcn}, as well as the emergence of open-access repositories of real networks, such as SNAP \cite{snapnets}, DBLP, or AMiner datasets.
There is thus a growing need for theoretical network models, as they provide solid grounds for statistical reasoning and allow for simplifying assumptions about data.
Citation graphs are among the most prominently analysed and modelled networks: the publications represent vertices and bibliographic references are stored as directed edges.
Such graphs help uncover patterns of knowledge dissemination, academic influence, and research trends over time~\cite{newman-networks} that are crucial topics in bibliometrics and science of science~\cite{Milojevic2025}.

In our work, we are specifically interested in modelling real networks as iteratively growing random graphs with a~community (cluster, subgroup) structure. The communities play a significant role in many types of real graphs, such as biological, urban~\cite{fortunato-survey}, and social networks~\cite{chen2014}, but their role in modern modelling of citation networks, where there is a division to scientific fields and knowledge domains, is often underplayed. Also, the problem of community detection has for a long time been prevalent in the area of network science~\cite{fortunato-survey}. While the problem of finding communities is long-standing and we already have access to highly effective methods of finding communities, such as the Louvain~\cite{blondel-louvain} and Leiden~\cite{traag-leiden} algorithms, they cannot directly help us distinguish which network parameters govern the emergence of community structure and how the communities change over time.

As such, we wish to propose a new network model featuring a community structure with a focus on the flexibility and interpretability of the underlying parameters. Our model can be applied to any network with either exogenous or algorithmically found communities, as long as they promote associative connections; that is, we require that citations are more likely to happen between two works from the same community. Also, while our model is derived on the ground of paper-to-paper citation networks, we also consider its applications to, e.g., author-to-author networks. To formulate our model concisely and keep the number of parameters reasonable, we formulate our proposition as a minimal analytical extension of an existing 3DSI model~\cite{3dsi-pnas} which we describe below.
Our generalisation is motivated by the necessity to model the heterogeneity inherent in scientific landscapes. Existing research indicates significant differences between how various scientific disciplines (represented here as communities or clusters) grow, characterised by distinct growth ratios, average reference list lengths, and preferential citing tendencies~\cite{Cascarina2023, Harzing2010, simkin}. A single global parameter set cannot capture these local dynamics; for instance, a citation in a~fast-moving field like Artificial Intelligence has different dynamical characteristics than one in Classical Logic.

Continuing, we shortly describe the core areas in literature related to our model: the foundational models of citation network science, the role of communities in networks, and key modern approaches to citation network modelling.

\paragraph{Foundational and citation-specific growth models.}
The earliest attempts to explain the scale-free nature of citation distributions focused on the ``rich-get-richer'' phenomenon. Price's model of cumulative advantage~\cite{deSollaPrice1965} was among the first to formalise preferential attachment, a concept which later inspired the Barabási–Albert model~\cite{BA-model}, and which highlighted the ubiquity of power-law distributions in complex networks.
However, simple preferential attachment inevitably leads to a strong bias towards older nodes. To address this, fitness-based models were introduced, such as the Bianconi–Barabási model~\cite{BB-model}, which assigns an intrinsic ``fitness'' to nodes to explain how younger papers can overtake older ones (e.g., ``sleeping beauties'').
More recently, specific attention has been paid to the temporal dynamics of citations. Models proposed by \citet{medo2011} and \citet{golosovsky2017} explicitly incorporate mechanisms for node relevance decay, demonstrating that the probability of receiving a citation must diminish over time to match empirical observations. This aligns with bibliometric studies on ``attention decay''~\cite{parolo2015} and the long-term impact of scientific works~\cite{wang2013}, which show that citation rates follow a rise-and-fall pattern not fully captured by standard scale-free models.

\paragraph{Modelling communities and structure.}
While growth models focus on temporal dynamics, a parallel stream of research focuses on the modular structure of networks. The Stochastic Block Model (SBM)~\cite{holland-sbm} serves as the generative baseline for networks with latent group structures. Unlike growth models, SBMs are typically static, but they offer rigorous connections between model parameters and structural properties; notably, \citet{decelle2011inference} identified precise phase transitions in SBMs that dictate when community structure becomes theoretically detectable.
Other approaches for generating clustered networks include the LFR Benchmark~\cite{lancichinetti-lfr} and GSCALER~\cite{gscaler}. While these tools are invaluable for benchmarking community detection algorithms, they are designed primarily to reproduce static structural features (e.g., degree distributions and community sizes) rather than to model the process of network growth via citation dynamics.
Topic models, such as Latent Dirichlet Allocation (LDA)~\cite{lda-model} and the Relational Topic Model (RTM)~\cite{rtm-model}, bridge this gap by representing papers as vectors of topic proportions and predicting links based on the similarity of these distributions.
However, these approaches often lack the analytical simplicity of pure random graph models. Notably, Drobyshevskiy and Turdakov~\citeyearpar{Drobyshevskiy-survey} published a comprehensive survey including eight random graph models capable of generating networks with a community structure, yet they found that none allowed for community structure under the preferential attachment rule. This highlights a critical gap in analytical models that simultaneously capture the \textit{temporal growth} of citation networks and their \textit{heterogeneous community structure}.

\paragraph{Recent approaches to citation network modelling.}
Many modern approaches generalise the previously mentioned core models or focus on data-driven generation of new graphs. Here we can mention Graph Neural Networks~\cite{kipfgcn} or Graph Variational Autoencoders~\cite{gvae}. However, these models often lack statistical precision and elegance and are based on heuristic assumptions.
Another approach gaining popularity is comprised of Exponential Random Graph Models (ERGMs), which can be viewed as generalisations of SBMs~\cite{fronczak-sbm}. Using a statistical framework, they define a probability distribution over the space of all possible graphs of a~certain size. While highly efficient in testing hypotheses on the networks, they still require computational-heavy fitting to real data.
On the other end of the spectrum are models such as the 3DSI (three dimensions of scientific impact) model~\cite{3dsi-pnas}; see Section~\ref{sec:3dsi}. It builds on the standard procedure of iterative preferential gains of the Barabási–Albert and Price models by making the split between accidental and preferential attachment explicit and parameterising it directly. The name comes from the model's ability to describe citation records using only three parameters: productivity, total impact, and how lucky an author has been so far. It is easy to fit to the data and offers valuable analytical insight into the data, but in its standard formulation it does not allow communities in the data.

\medskip
We thus lack tools for analytically modelling the citation networks with communities via the lens of principled growth-based models. There are analytical models for simpler networks (BA, Price, 3DSI) and technically complex models suitable for communities (GNNs, GVAs; see also~\citealp{tian2023}), but these on the other hand offer no theoretical insight and require black-box explanations. In the current contribution, we would like to offer a solution to these issues: our proposal, which we set forth in Section~\ref{sec:model}, is a generalisation of the 3DSI model which distinguishes between two types of citations: accidental, which are purely random, and preferential, utilising the rich-get-richer scheme. While the 3DSI model was shown to fit quite well to some real networks~\cite{pricepareto2}, it is based on the assumption of relative homogeneity of the papers in general, and as such is not ideal for modelling networks with local variability, such as networks with a~community structure.

We prove some theoretical results regarding, amongst others, the Gini index
of the node degree distribution, as well as some asymptotic properties, such as the limiting distribution's being Paretian (Section~\ref{sec:largen}). In Section~\ref{sec:experiments}, we demonstrate
the new model's usefulness in describing real-world citation networks.
Section~\ref{sec:conclusion} concludes the paper.

\section{New model}\label{sec:model}

\subsection{Homogeneous Price–Pareto model}\label{sec:3dsi}

Here are the basic assumptions behind the standard 3DSI citation network growth presented in~\cite{3dsi-pnas}.
\begin{itemize}
    \item In each iteration, a new publication is added to an existing citation network.
    \item The new work distributes $m$ references among the nodes already in the network, such that for some parameter $\rho$:
    \begin{itemize}
        \item $(1-\rho)m$ citations are allocated uniformly at random,
        \item after that, $\rho m$ citations are allocated based on preferential attachment, i.e., nodes are selected with probabilities proportional to their degrees.
    \end{itemize}
    \item The above procedure can be repeated indefinitely or until a desired number of vertices has been created.
\end{itemize}

Throughout the paper, all gains are considered in expectation. In particular, values such as $(1-\rho)m$ need not be integer. \citet{pricepareto2} showed that the approximate degree $d^{(t)}(\ell)$ of a node $\ell$ at time $t$ follows the recurrence relation:
\[
\underbrace{d^{(t)}(\ell)}_{\substack{\text{the degree of node $\ell$} \\ \text{in turn $t$}}}= \underbrace{d^{(t-1)}(\ell)}_{\substack{\text{the degree of node $\ell$} \\ \text{in turn $t-1$}}} + \underbrace{\Acc{}^{(t)}(\ell)}_{\substack{\text{accidental citation gain}}} + \underbrace{\rho m\frac{d^{(t-1)}(\ell) + \Acc^{(t)}(\ell)}{(t-1)m + (1-\rho)m}}_{\text{preferential citation gain}}.
\]
We can instantly calculate $\Acc^{(t)}(\ell) = \frac{(1-\rho)m}{t}$ as the number of citations added accidentally divided equally by all $t$ nodes present at time $t$ (note $t$ and not $t-1$ since 3DSI allows loops). This formula, with the starting condition $d^{(\ell)}(\ell)=0$, resolves to an analytical equation:
\begin{equation}
\label{eq:standard-3dsi}
d^{(t)}(\ell) = \frac{m(1-\rho)}{\rho}\left(\frac{\Gamma(\ell - \rho)\Gamma(t+1)}{\Gamma(\ell)\Gamma(t+1-\rho)}-1\right).
\end{equation}

Conceptually, 3DSI is closely related to the Price \citeyearpar{deSollaPrice1965} power-law-type model.
In the latter, the preferential citation gain of node $\ell$ is proportional to $d_i^{(t)}(\ell) + C$ for some constant $C$, so that papers with no citations can still attract new ones.
The 3DSI model achieves the same effect through a different but equivalent mechanism. Namely, in each turn, the accidental citations are added first. Only afterwards the preferential ones are incorporated. This way, every article has a non-zero expected in-degree before the preferential attachment comes into play. The additive constant $C$ and the accidental rate $1-\rho$ are equivalent in that they account for the accidental fraction of the citations~\cite{newman-networks}, though the two differ in interpretation. Moreover, as we will mention below, asymptotically, 3DSI enjoys convergence to a Pareto type II distribution. To account for its standing on the shoulders of giants, we will also be referring to 3DSI as the Price--Pareto model.

\subsection{Introducing community structure}

Even though the original model was empirically shown to fit well to some existing citation networks on a large portion of the domain, its standard formulation does not naturally allow for the emergence of communities, especially not ones with a priori ground-truth information, which is particularly useful in community detection tasks. Therefore, we propose to extend it in the following way:

\begin{enumerate}
    \item In each iteration, a new node is assigned the community $i$ with probability $p_i$, such that there are $k$ communities in total and $\sum_{i=1}^k p_i=1$.
    \item Both parameters $m$ and $\rho$ are now community-specific: they depend on the cluster the new node belongs to; the $i$-th cluster has its own $m_i$ and $\rho_i$, respectively.
    \item When allocating accidental edges, the citation-receiving paper is selected uniformly at random from the whole network. Preferential edges are drawn only from the vertices from the same community as the new node. The loops are no longer allowed.
\end{enumerate}
Such an algorithm naturally extends the standard 3DSI formulation based on empirically observed phenomena:
\begin{itemize}
    \item papers from different scientific domains tend to cite papers from own domains in different ratios; see, e.g., \cite{Cascarina2023} and \cite{Harzing2010},
    \item different domains gain articles with different rates; see \cite{OlejniczakSavageWheeler2022},
    \item papers from various domains have varying average reference lists' lengths; see \cite{Dai2021}.
\end{itemize}

Moreover, time is now considered \emph{locally}, i.e., with respect to the current size of a selected community $i$. Since the communities are drawn randomly, we expect that the community $i$ grows by one paper at a time when the total network grows by approximately $\frac{1}{p_i}$ papers.

\paragraph{Notation.}
In the following derivations, let $t$ denote the discrete local time step in the $i$-th community ($t_i$ suppressing the subscript), corresponding directly to the number of vertices currently assigned to that community.
Let $d_i^{(t)}(\ell)$ denote the in-degree of a vertex $\ell$ belonging to community $i$ at local time $t$.
Finally, let $\Acc_i^{(t)}(\ell)$ denote the expected accidental gain accumulated by vertex $\ell$ in community $i$ during the interval between local time steps $t-1$ and $t$. In the derivations, we frequently make use of the fact that the vertex identifier corresponds to its entry time.

\paragraph{Iterative formula.}
We can formulate a general recurrence relation:
\begin{equation}
\label{eq:general-formula}
\underbrace{d_i^{(t)}(\ell)}_{\substack{\text{the in-degree of node $\ell$ in cluster $i$} \\ \text{in (local) turn $t$}}} = \underbrace{d^{(t-1)}_i(\ell)}_{\substack{\text{the in-degree} \\ \text{in the previous turn}}} + \underbrace{\Acc{}_i^{(t)}(\ell)}_\text{accidental citation gain} + \underbrace{\rho_i m_i \frac{d_i^{(t-1)}(\ell) + \Acc_i^{(t)}(\ell)}{\sum_{r=1}^{t-1}\left[d_i^{(t-1)}(r)+\Acc^{(t-1)}_i(r)\right]}}_\text{preferential citation gain}.
\end{equation}
To calculate $\Acc_i^{(t)}(\ell)$, we make a few observations:
\begin{enumerate}
    \item At the local turn $t-1$, we know that there are exactly $t-1$ vertices in the community $i$. Unfortunately, we do not know the exact size of the whole network:
    let us temporarily denote it with $X$. Note that  $X$ represents the total number of turns it took to sample exactly $t-1$ vertices from the class $i$ with probability $p_i$.  We can thus model $X$ as coming from the negative binomial\footnote{There are notational differences for the negative binomial distribution in the literature. We use the version where $X$ represents the total number of trials required for $t-1$ successes, which implies that the support of $X$ is $[t-1, \infty) \cap \mathbb{N}$.} distribution $X \sim \negbin(r = t-1, p = p_i)$, i.e., the probability mass function of $X$ is given by the formula:
    \begin{equation*}
        \mathbb{P}(X=x)= {\binom {x-1}{x-r}}(1-p)^{x-r}p^{r}.
    \end{equation*}

    \item Since the network is growing, we consider the accidental gain for the node $\ell$ at each of the global turns $X+s$ for some $s \geq 0$. The variable $s$ increases as long as we do not draw a new vertex from the community $i$. So
    $(1-p_i)^{s-1}$ is the probability that the number of global steps between the local time $t-1$ and $t$ is equal to at least $s$.
    \item Since the community probabilities are independent of $X$ and $t$, we can write that the expected accidental gain of the vertex $\ell$ in the global turn $X+s$ is equal to:
    \[
    \sum_{j=1}^k p_j \frac{m_j(1-\rho_j)}{X+s} = \frac{\langle m \rangle - \langle \rho m \rangle}{X+s},
    \]
    where $\langle m \rangle = \sum_{j=1}^k p_j m_j$ and $\langle \rho m \rangle = \sum_{j=1}^k p_j m_j \rho_j$. Note that $\langle m \rangle - \langle \rho m\rangle$ is a weighted average number of accidental citations gained. For brevity, we further denote it by $\langle a \rangle := \langle m \rangle - \langle \rho m \rangle$.

    \item What follows is that the total expected accidental gain of the node $\ell$ in the local turn $t-1$ is equal to:
    \[
    \Acc{}^{(t)}_i(\ell) =\mathbb{E}_X\left[ \sum_{s=0}^\infty (1-p_i)^{s} \frac{\mrm}{X+s} \right] = \mrm\sum_{s=0}^\infty (1-p_i)^s \mathbb{E}_X\left[\frac{1}{X+s}\right].
    \]

    \item From the properties of the Beta function $B$, we have that for positive $X+s$:
    \[
    B(1, X+s) = \frac{1}{X+s} = \int_0^1 u^{X+s-1} du,
    \]
    we apply Fubini's theorem and transform:
    \begin{align}
    \label{eq:accidental-part}
    \Acc{}_i^{(t)}(\ell) &= \mrm\sum_{s=0}^\infty (1-p_i)^s \int_0^1 \mathbb{E}_X\left[u^{X+s-1}\right] du = \notag \\
    &= \mrm\int_0^1 \frac{\mathbb{E}_X\left[u^{X-1}\right]}{1 - (1-p_i)u} du.
    \end{align}

 \item Since $X$ follows the negative binomial distribution, we can use the formula for the probability generating function of $X$ as:
    \[
\mathbb{E}_X\left[u^{X}\right] = \left(\frac{p_iu}{(1-(1-p_i)u)}\right)^{t-1} \implies \mathbb{E}_X\left[u^{X-1}\right] = \frac{p_i^{t-1} u^{t-2}}{(1-(1-p_i)u)^{t-1}},
    \]
and thus:

        \begin{align}
    \label{eq:accidental-part2}
    \Acc{}_i^{(t)}(\ell) &=    \mrm\int_0^1 \frac{p_i^{t-1} u^{t-2}}{(1-(1-p_i)u)^{t}} du = \notag \\
    &= \frac{\mrm}{t-1}.
    \end{align}

\end{enumerate}
Before applying Formula~\eqref{eq:accidental-part2} to Equation~\eqref{eq:general-formula}, let us also note that the total in-degree sum of cluster $i$ after the local turn $t$, denoted by $\Sigma_i^{(t)}$, approximately follows:
\begin{equation}
\label{eq:sigma}
    \Sigma_i^{(t)} \approx \Sigma_i^{(t-1)} + \rho_im_i + \sum_{r=1}^{t-1} \Acc_i^{(t)}(r) = \Sigma_i^{(t-1)} + \rho_i m_i + \mrm = (\mrm + \rho_i m_i)(t-1),
\end{equation}
where we assume $\Sigma_i^{(1)} = 0$, which is natural as it disallows self-loops in the graph.
Finally, applying both results to Equation~(\ref{eq:general-formula}), we get:

\begin{align*}
    d^{(t)}_{i}(\ell) &= d^{(t-1)}_{i}(\ell) + \frac{\mrm}{t - 1} +  \rho_i m_i \frac{d^{(t-1)}_{i}(\ell) + \frac{\mrm}{t-1}}{\sum_{r=1}^{t-1} \left[ d^{(t-1)}_{i}(r) + \frac{\mrm}{t-1} \right] }=\\
    &=\left[d^{(t-1)}_{i}(\ell) + \frac{\mrm}{t - 1}\right]  \frac{t-1}{t-1- \nu_i },
\end{align*}
where we introduce $\nu_i$ as:
\begin{equation}
\label{eq:nu}
\nu_i=1-\frac{\mrm}{\mrm+\rho_i m_i }=\frac{\rho_i m_i}{\mrm+\rho_i m_i}.
\end{equation}
In the standard version of 3DSI, we have $\rho = \frac{\rho m}{m} = \frac{\rho m}{(1-\rho)m + \rho m}$. As such, $\nu_i$ can be considered an ``effective'' version of $\rho$ from the original model. Assuming that $p_i \in (0, 1)$, $\rho_i \in (0, 1)$, $k>0$, and $m_i > 0$,
we have that
$\nu_i \in (0, 1)$.

\paragraph{Closed-form formula.}
Solving the recurrence using the telescopic rule, similarly to the solution presented in~\cite{3dsi-pnas}, and setting the starting condition $d_i^{(\ell)}(\ell)=0$, we obtain:
\begin{equation}
\label{eq:analytical-formula}
d^{(t)}_{i}(\ell) =\frac{\mrm}{\nu_i}\left[
\frac{\Gamma(\ell-\nu_i)}{\Gamma(\ell)}
\frac{\Gamma(t)}{\Gamma(t-\nu_i)}
-1
\right].
\end{equation}

\begin{figure}
    \centering
    \includegraphics[width=1\linewidth]{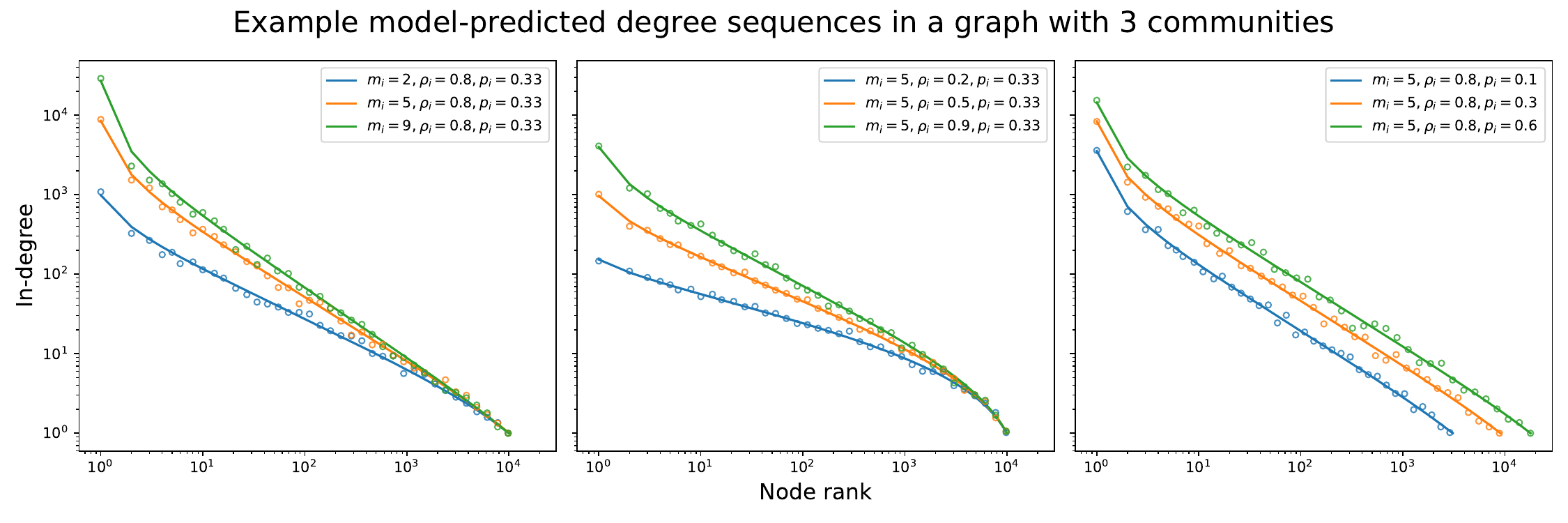}
    \caption{Example model-predicted node in-degree sequences for communities in a graph with $N=30{,}000$ nodes and three communities. Symbols show the average in-degrees over $50$ independent simulations of the described process.}
    \label{fig:example-degrees}
\end{figure}

\paragraph{Sum of in-degrees.}
We can check that the in-degrees have the correct sum:
\begin{align*}
\sum_{\ell=1}^t d_i^{(t)}(\ell) &= \frac{\mrm}{\nu_i }\left(-t+\sum_{\ell=1}^t \frac{\Gamma(t)\Gamma(\ell-\nu_i)}{\Gamma(\ell)\Gamma(t-\nu_i)}\right)= \\
& = \frac{\mrm}{\nu_i} \left(-t + \frac{t-\nu_i}{1-\nu_i}\right) \stackrel{\text{Eq.\eqref{eq:nu}}}{=} \left(\mrm+\rho_i m_i\right)(t-1).
\end{align*}
This confirms that the assumption~\eqref{eq:sigma} remains satisfied during the network growth.

It is tempting to try convert the formula~\eqref{eq:analytical-formula} to a formula of a node in-degree in the global sequence. The degree of the vertex $\ell$, aside from its community probability $p_i$, depends also on both the number of vertices from the same cluster that precede it and succeed it in the global sequence. Combining with the fact that the values of $\nu_i$ are not independent between clusters, unfortunately, we found no simple analytical representation of the global degree sequence in this model.

Figure~\ref{fig:example-degrees} presents the node in-degrees output by our model for a graph with three communities on $30{,}000$ vertices. We observe that modifying different parameters has impact on different aspects of the final sequence: higher $m_i$ mostly impacts the overall slope of the sequence, $\rho_i$ impacts the curvature of the degree sequence the most, while $p_i$  changes the amplitude of the sequence as it directly represents the number of vertices in the community.

\subsection{Gini index}

Noting that the above in-degrees are sorted decreasingly, we can also obtain the Gini coefficient of the vertex degrees in the $i$-th cluster in local time $t$, denoted by $\mathcal{G}_i^{(t)}$:
\begin{align}\label{eq:gini}
    \mathcal{G}_i^{(t)}&=    \frac{1}{(t-1)\Sigma_i^{(t)}} \sum_{\ell=1}^t (t-2\ell+1) d_i^{(t)}(\ell)=\notag\\
    &=\frac{1}{(t-1)\Sigma_i^{(t)}}\frac{\mrm}{\nu_i} \left[(t+1)\sum_{\ell=1}^t  \frac{\Gamma(t)\Gamma(\ell-\nu_i)}{\Gamma(\ell)\Gamma(t-\nu_i)}-2\sum_{\ell=1}^t \ell \frac{\Gamma(t)\Gamma(\ell-\nu_i)}{\Gamma(\ell)\Gamma(t-\nu_i)}\right]=\notag\\
     &=\frac{\mrm}{\Sigma_i^{(t)}} \frac{t-\nu_i}{(1-\nu_i)(2-\nu_i)}= \frac{t-\nu_i}{t-1} \frac{1}{2-\nu_i},
\end{align}
where the last equality follows by applying Equation~\eqref{eq:sigma}.
Of course, for large $t$,
$\mathcal{G}_i^{(t)}\approx \frac{1}{2-\nu_i}$.

As with the global degree sequence, finding the Gini coefficient of the whole sequence is difficult due to the mixture nature of the modelled sample. We provide clean asymptotic results though, which we derive in the next section.

\subsection{Asymptotic results: Price meets Pareto}\label{sec:largen}

We can consider the theoretical formulae asymptotically for $t \rightarrow \infty$, as both network sizes and community sizes are often sufficiently big. Following~\cite{pricepareto2}, we consider the limit of the in-degree in terms of the survival function:
\begin{equation}
    \lim_{t \rightarrow \infty} d_i^{(t)}(yt) = \lim_{t \rightarrow \infty} \frac{\mrm}{\nu_i }\left[\frac{\Gamma(t)\,\Gamma(yt-\nu_i)}{\Gamma(yt)\,\Gamma(t-\nu_i)}-1 \right], %
\end{equation}
where $y \in [0, 1]$ is the normalised rank. Since $\nu_i \in (0, 1) \implies 1-\nu_i \in (0, 1)$, from Gautschi inequality~\cite{Gautschi1959} we have:
\[
\left(\frac{t-1}{yt}\right)^{\nu_i}<\frac{\Gamma((t-1)+1)\,\Gamma((yt-1)+(1-\nu_i))}{\Gamma((t-1)+(1-\nu_i))\,\Gamma((yt-1)+1)} < \left(\frac{t}{yt-1}\right)^{\nu_i}.
\]
Since the limit of both the lower and the upper bounds as $t \rightarrow \infty$ is equal to $y^{-\nu_i}$ and denoting $S^{-1}_i(y)$ as the inverse of the survival function, we get:
\[
S_i^{-1}(y)=\lim_{t \rightarrow \infty} d_i^{(t)}(yt)=\frac{\mrm}{\nu_i} \left(y^{-\nu_i}-1\right).
\]
Writing $x := S_i^{-1}(y)$ for this limiting in-degree and inverting:
\[
S_i (x) = \left(\frac{\nu_i}{\mrm}x+1\right)^\frac{-1}{\nu_i}.
\]
Note that $F_i (x) := 1 - S_i(x)$ may be considered the CDF of the in-degree distribution in cluster $i$ for sufficiently large $t$. We obtain that the density of the degree distribution can thus be modelled as:
\[
f_i (x) = \frac{dF_i(x)}{dx} = \frac{1}{\mrm} \left(1 + \frac{\nu_i }{\mrm}x\right)^{-1-\frac{1}{\nu_i}},
\]
which is the Pareto type II (Lomax) distribution \cite{arnold2015pareto} with parameters $\alpha = \frac{1}{\nu_i}$ and $\lambda=\frac{\mrm}{\nu_i}$; compare \cite{OurGiniLorenz} for a similar result regarding the standard 3DSI model.

As a side note, from this it directly follows that the asymptotic expected degree exists if $m_i \rho_i p_i > 0 \implies \nu_i > 0$; then $\mathbb{E}(X)_{X \sim f_i} = \frac{\mrm}{1-\nu_i}$. Indeed:
\[
\lim_{t \rightarrow \infty} \sum_{\ell=1}^t \frac{1}{t} d^{(t)}(\ell) \stackrel{\text{Eq.\eqref{eq:sigma}}}{=} \lim_{t \rightarrow\infty}\frac{t-1}{t} \left(\mrm+\rho_im_i\right) = \mrm+\rho_im_i = \frac{\mrm}{1-\nu_i} =: \mu_i.
\]

\paragraph{Overall Gini index.}
Since each paper comes from a cluster $i$ with probability $p_i$, we have that the in-degree of a randomly selected paper comes from a mixture distribution $F(x) = \sum_{j=1}^k p_i F_i(x)$. Without loss of generality, assume that for all $i \leq j$, $\nu_i \leq \nu_j$.
By \cite{gini-formulas}, this allows us to approximate the Gini coefficient $\mathcal{G}$ of the whole degree sequence:
\begin{align}\label{eq:totalgini}
    \mathcal{G} &= \frac{1}{\langle \mu \rangle}\int_{0}^\infty S(x)(1-S(x)) dx =\\
    &= \frac{1}{\langle \mu \rangle} \sum_{i=1}^k p_i \int_0^\infty S_i(x) dx - \frac{1}{\langle \mu \rangle} \sum_{i, j = 1}^k p_i p_j \int_0^\infty S_i(x) S_j(x) dx,
\end{align}
 where $\langle \mu \rangle = \sum_{i=1}^k p_i \mu_i$. Let us also denote:
 \[
 A_i := \frac{\mrm}{\nu_i}.
 \]
 Then $S_i(x) = \left(\frac{x}{A_i}+1\right)^\frac{-1}{\nu_i} = A_i^\frac{1}{\nu_i} (x+A_i)^\frac{-1}{\nu_i}
 $ and from our assumptions follow that $\frac{1}{\nu_i} > 1, A_i > 0$. We directly obtain:
 \begin{equation}\label{eq:survival1}
 \int_0^\infty S_i(x) dx = A_i^\frac{1}{\nu_i} \int_0^\infty (x+A_i)^\frac{-1}{\nu_i} dx = A_i^\frac{1}{\nu_i} \frac{A_i^{1-\frac{1}{\nu_i}}\nu_i}{1-\nu_i} = \frac{A_i\nu_i}{1-\nu_i} = \frac{\mrm}{1-\nu_i} = \mu_i,
 \end{equation}
and
\begin{align}\label{eq:gini-pair}
    R_{i, j} := \int_0^\infty S_i(x) S_j(x) dx = A_i^\frac{1}{\nu_i}A_j^\frac{1}{\nu_j}\int_0^\infty (x+A_i)^\frac{-1}{\nu_i}(x+A_j)^\frac{-1}{\nu_j}dx.
\end{align}
Following, e.g., Formula 3.197.1 from~\cite{gradshtein-definite-integrals} we get:
\begin{align}
R_{i, j} &= A_i^\frac{1}{\nu_i}A_j^\frac{1}{\nu_j} A_i^\frac{-1}{\nu_i}A_j^{1-\frac{1}{\nu_j}} B\left(1, \frac{1}{\nu_i}+\frac{1}{\nu_j}-1\right) {}_{2}{F}_1\left(\frac{1}{\nu_i}, 1; \frac{1}{\nu_i}+\frac{1}{\nu_j}; 1-\frac{A_j}{A_i}\right) \notag \\
&= A_j \frac{\nu_i \nu_j}{\nu_i + \nu_j - \nu_i \nu_j}{}_{2}{F}_1\left(\frac{1}{\nu_i}, 1; \frac{1}{\nu_i}+\frac{1}{\nu_j}; 1-\frac{\nu_i}{\nu_j}\right),
\end{align}
where $B(x, y)$ is the Beta function and ${}_{2}{F}_1(a, b;c;z)$ is the hypergeometric function. Note that we use the assumption that $\nu_i \leq \nu_j$ to make sure the hypergeometric series is convergent due to $1-\frac{\nu_i}{\nu_j} < 1$.  For $i=j$, the hypergeometric function simplifies to:
\[
R_{i, i} = A_i \frac{\nu_i^2}{2\nu_i - \nu_i^2}{}_{2}F_1\left(\frac{1}{\nu_i}, 1; \frac{2}{\nu_i}; 0\right) = \frac{A_i \nu_i}{2 - \nu_i} = \frac{\mrm}{2-\nu_i},
\]
which is also consistent with the formula used to evaluate Equation~\eqref{eq:survival1} if we applied it to $\int_0^\infty S^2_i(x)dx$. We could also further approximate the hypergeometric function:
\begin{align}
    {}_{2}F_1\left(\frac{1}{\nu_i}, 1; \frac{1}{\nu_i}+\frac{1}{\nu_j}; 1 - \frac{\nu_i}{\nu_j}\right) = \sum_{n=0}^\infty \frac{\Gamma(\frac{1}{\nu_i}+n)\Gamma(\frac{1}{\nu_i}+\frac{1}{\nu_j})}{\Gamma(\frac{1}{\nu_i})\Gamma(\frac{1}{\nu_i}+\frac{1}{\nu_j}+n)} \left(\frac{\nu_j - \nu_i}{\nu_j}\right)^n \approx \sum_{n=0}^\infty \frac{1}{\nu_i^n} \left(\frac{1}{\nu_i}+\frac{1}{\nu_j}\right)^{-n} \left(\frac{\nu_j - \nu_i}{\nu_j}\right)^n,
\end{align}
where the approximation comes from taking the first terms of the Stirling approximation of the factorial. Tidying up the terms and applying the formula for the sum of the geometric series, we find:
\begin{equation}\label{eq:2f1approx}
   {}_{2}F_{1}\left(\frac{1}{\nu_i}, 1; \frac{1}{\nu_i}+\frac{1}{\nu_j}; 1-\frac{ \nu_i}{ \nu_j} \right) \approx \frac{\nu_i + \nu_j}{2\nu_i}.
\end{equation}
Finally, we can check that the expected in-degree is correct:
\begin{equation}
\label{eq:mucheck}
\langle \mu \rangle = \sum_{i=1}^k p_i \mu_i = \mrm \sum_{i=1}^k \frac{1}{1-\nu_i} = \mrm \sum_{i=1}^k \frac{\mrm + \rho_i m_i}{\mrm} = \langle m \rangle.
\end{equation}
Overall, combining Equations \eqref{eq:totalgini}-\eqref{eq:mucheck} allows for a simpler representation of the Gini coefficient of the whole sample:
\begin{align}
\mathcal{G} &= 1 - \frac{1}{\langle m \rangle}\sum_{i, j =1}^k p_i p_j R_{i, j} = 1 - \frac{\mrm}{\langle m \rangle}\sum_{i, j=1}^k p_i p_j \frac{\nu_i }{\nu_i + \nu_j - \nu_i \nu_j}{}_2F_{1}\left(\frac{1}{\nu_i}, 1; \frac{1}{\nu_i}+\frac{1}{\nu_j}; 1-\frac{\nu_i}{ \nu_j} \right) \notag \\
&\approx 1 - \frac{\mrm}{\langle m \rangle} \sum_{i, j=1}^k p_i p_j \frac{\nu_i + \nu_j}{2(\nu_i + \nu_j - \nu_i\nu_j)}.
\end{align}
Numerical tests conducted on a battery of samples with various sets of parameters and $10{,}000 \leq N \leq 20{,}000$ indicate that the difference between the Gini coefficient value calculated with the exact formula and with the approximation almost never exceeds $0.02$ and that on average the difference is very close to $0$. While~\cite{pricepareto2} also presents clean derivation of the order statistics for the standard 3DSI degree sequence, we have found that introducing the community mixture makes the resulting formulae unsimplifiable (though they can be represented using, i.a., the Lauriciella generalised hypergeometric functions). As such, we omit this derivation here.

\section{Experiments}\label{sec:experiments}

\subsection{Datasets and data curation}

We fit our model to two well-established datasets. First, we utilise the Cora dataset \cite{cora-dataset}, a benchmark in machine learning research consisting of 2,708 scientific papers classified into seven distinct sub-disciplines (e.g., Neural Networks, Genetic Algorithms). This dataset represents a standard paper-to-paper citation network where edges are directed citations. It is widely used as a benchmark dataset in tasks such as node classification.

Second, we analyse the DBLP v14 dataset \cite{dblp-dataset}, a comprehensive bibliographic database of computer science publications. We use the fourteenth edition as it contains field-of-study information, which we use as the source for ground-truth communities. To study an author-to-author topology, we projected the paper-to-paper graph into a simple directed graph where nodes represent paper authors. Specifically, a directed edge is created from author $u$ to author $v$ if author $u$ has written a paper that cites any paper written by author $v$. To adhere to our model's assumptions (simple directed graph), we flattened the resulting multigraph by treating multiple citations between the same pair of authors as a single unweighted edge and removed self-loops (self-citations). The ground-truth community of each author was determined by weighted majority vote: the author is assigned to the community corresponding to the discipline with the highest total weight over all the papers of this author, where the weight was extracted from the field-of-study information in the data. The motivation for the transformation we performed on DBLP is two-fold: first, it allows us to reduce the noise in ground-truth community labels that were assigned automatically; second, it allows us to test our model on an author-to-author network instead of a standard paper-to-paper network.

\begin{table}[]
    \centering
    \caption{General information about the analysed real networks}
    \begin{tabular}{l|r|r|r|r}
    Dataset & Number of nodes & Number of edges & Number of communities & Average in-/out-degree \\ \hline
    CORA & 2{,}708 & 5{,}429 & 7 & 2.005\\
    DBLP V14 Authors & 481{,}386 & 57{,}781{,}812 & 8 & 120.032
    \end{tabular}

    \label{tab:datasets}
\end{table}

\begin{figure}
    \centering
    \includegraphics[width=1\linewidth]{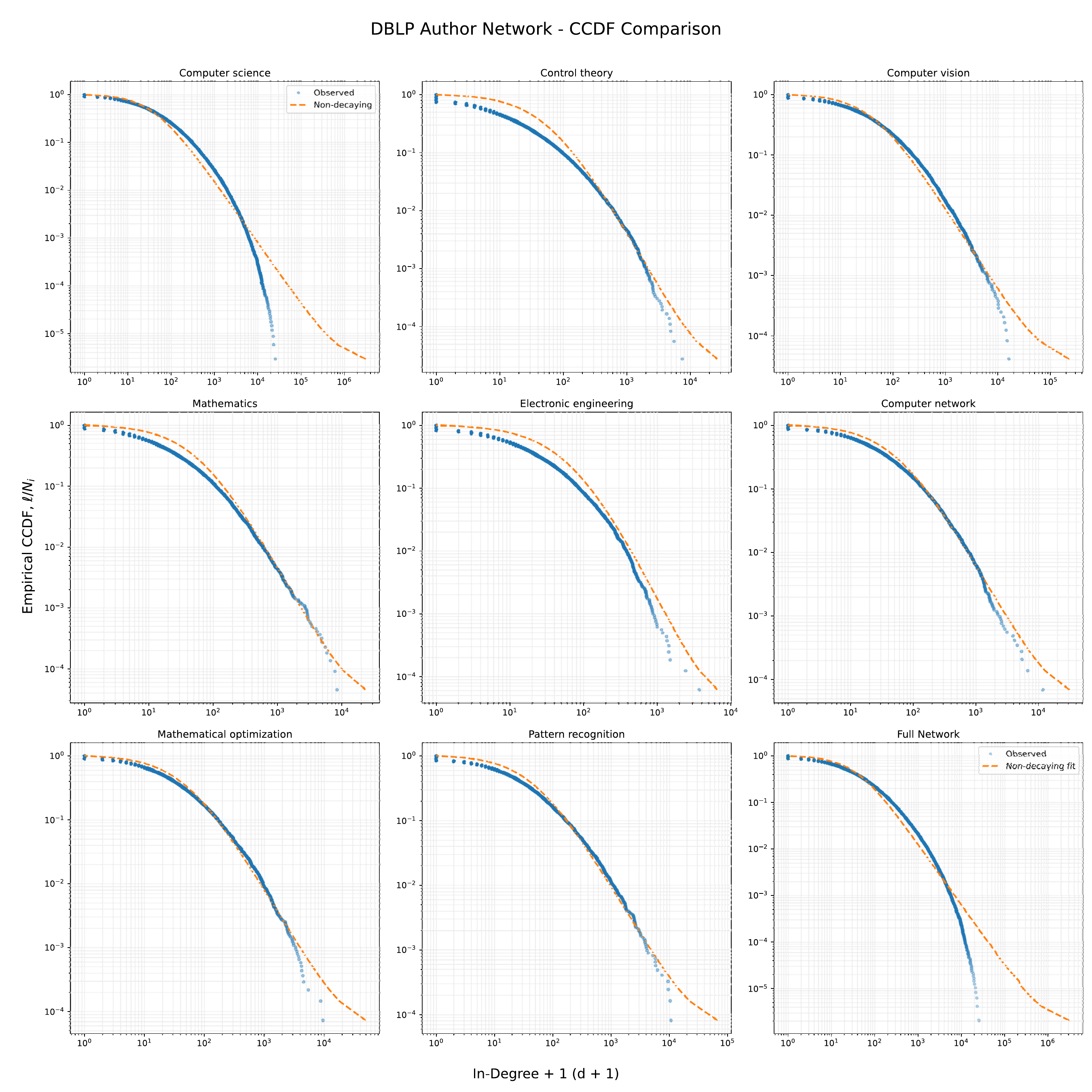}
    \caption{Complementary cumulative distribution functions of the observed and modelled in-degree sequences in the eight clusters of the DBLP authors dataset. Our model was fitted using the Gini index-based $\rho_i$ estimators. The last plot shows the full in-degree sequences.}

    \label{fig:dblp-density}
\end{figure}
\begin{figure}
    \centering
    \includegraphics[width=1\linewidth]{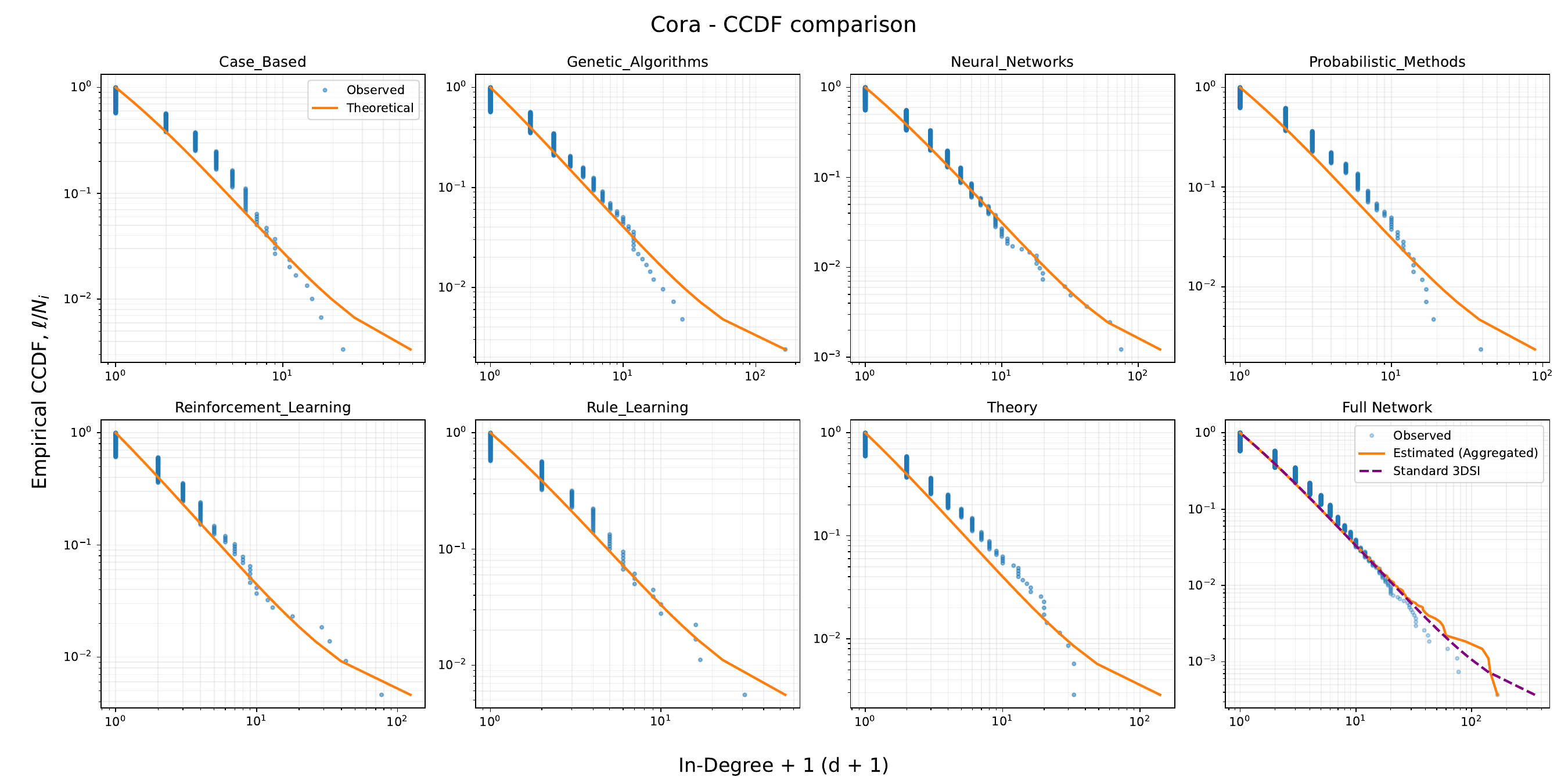}
    \caption{Complementary cumulative distribution functions of the observed and modelled in-degree sequences in the seven clusters of the Cora dataset. Our model was fitted using the Gini index-based $\rho_i$ estimators. The last plot shows the full in-degree sequences.}
    \label{fig:cora-density}
\end{figure}

\subsection{Estimating network parameters}

To fit our model to a real network given as an unweighted directed graph, we can estimate the parameters in a straightforward manner from the graph's adjacency matrix. After computing:
\begin{enumerate}
\item $\Sigma_i$ -- the sum of in-degrees in the cluster $i$,
\item $\Psi_i$ -- the sum of the out-degrees in the cluster $i$,
\item $N_i$ -- the number of vertices in the cluster $i$,
\item $\mathcal{G}_i$ -- the Gini coefficient of the in-degree sequence in the cluster $i$,
\item $N=\sum_{i=1}^k N_i$ -- the total number of papers,
\end{enumerate}
we obtain:
\begin{equation}\label{eq:parameters}
\hat{m}_i = \frac{\Psi_i}{N_i-1}, \quad \hat{p}_i = \frac{N_i}{N} \quad \hat{\rho}_i = \frac{\Sigma_i (2 \mathcal{G}_i + N_i - 2\mathcal{G}_i N_i)}{\Psi_i (\mathcal{G}_i +1 - \mathcal{G}_i N_i)},
\end{equation}
and the last equation follows from solving Equation~\eqref{eq:gini} for $\rho_i$:
\[
\mathcal{G}_i = 1 - \frac{(N_i-2)\mrm}{(t-1)(m_i \rho_i + 2\mrm)} \implies \rho_i = \frac{\mrm(2\mathcal{G}_i+N - 2\mathcal{G}_iN_i)}{(\mathcal{G}_i-1)m_i (N_i-1)}.
\]
From Equation~\eqref{eq:sigma}, we have that $\mrm \approx \frac{\Sigma_i}{N_i-1}-\rho_i m_i$, so substituting:
\[
\rho_i = \left(\frac{\Sigma_i}{N_i-1}-\rho_i m_i\right) \frac{(2\mathcal{G}_i+N - 2\mathcal{G}_iN_i)}{(\mathcal{G}_i-1) m_i (N_i-1)} \implies \rho_i = \frac{\Sigma_i (2\mathcal{G}_i + N_i - 2\mathcal{G}_i N_i)}{m_i (N_i - 1)(1 + \mathcal{G}_i - \mathcal{G}_i N_i)}.
\]
Finally, by substituting $m_i \approx \hat{m}_i = \frac{\Psi_i}{N_i-1}$,
we obtain Formula~\eqref{eq:parameters}.

\bigskip
Aside from Formula~\eqref{eq:parameters}, there are also other ways to estimate $\rho_i$ in real networks if we assume that they follow our theoretical model. Below we present a few other possibilities.

\paragraph{Intra-cluster edges.} If we have access to the full network, we can calculate the total number of the intra-cluster citations in each community $i$, denoted by $\Psi_{i, i}$ (i.e., the sum of citations having both source and target in the community $i$). Note that each node, when it appears, gives out $\rho_i m_i$ preferential citations in intra-cluster edges and $(1-\rho_i)m_ip_i$ accidental intra-cluster citations. This yields:
\begin{equation}
\label{eq:rho-intra-cluster}
\Psi_{i, i} \approx (N_i-1)\left(\rho_i m_i + (1-\rho_i)m_ip_i\right) \implies \hat{\rho}_i = \frac{\frac{\Psi_{i, i}}{N_i-1}-\hat{m}_i\hat{p}_i}{\hat{m}_i(1-\hat{p}_i)} = \frac{N \Psi_{i, i} - N_i \Psi_i}{(N-N_i)\Psi_i}.
\end{equation}

\paragraph{In-degree sum equation.} Note that we can combine Equation~\eqref{eq:sigma} for all clusters to obtain a system of linear equations to be solved with respect to the vector $\mathbf{r} = [\rho_1, \rho_2, \dots, \rho_i]$. Assuming $\mathbf{A} = (\alpha_{i, j})_{k \times k}$ and $\mathbf{b} = (b_i)_k$, we get:
\[
\mathbf{A} \mathbf{r}^T = \mathbf{b},
\]
where:
\[
\alpha_{i, j} = \begin{cases}
    -p_jm_j &j \neq i,\\
    (1-p_j)m_j & j = i,
\end{cases}
\]
and:
\[
b_i = \frac{\Sigma_i^{(t)}}{t-1}-\langle m \rangle.
\]
The matrix $\mathbf{A}$ is singular with rank at most $k-1$ since
the average in-degree is also the average out-degree.
After plugging-in the estimates for $m_i, t$, we get:
\begin{equation}
\label{eq:sum-based-rho}
  \mathbf{r} = \frac{\Sigma_i^{(t)} - \langle m \rangle (N_i-1)}{\Psi_i} + x\left(\frac{N_1-1}{\Psi_1}, \frac{N_2-1}{\Psi_2}, \dots, \frac{N_k-1}{\Psi_k}\right)^T,
\end{equation}
where the first term is the same regardless of the chosen $i$, and $x \in \mathbb{R}$ is a free parameter. To ensure that each $0 < \rho_i < 1$, we need to have:
\[
x \in \left(\max_i\left(-b_i\right),  \ \min_i \left(m_i - b_i\right)\right)
\]
There is a number of ways to choose $x$ in the estimation: we could select a community and require any of the previous estimates (Equations~\eqref{eq:parameters} and \eqref{eq:rho-intra-cluster}) to hold for that community. Alternatively, we could fit $x$ to the data numerically.

Our qualitative analysis showed that estimators~\eqref{eq:rho-intra-cluster} and~\eqref{eq:sum-based-rho}, while analytically interesting, are subpar to the estimator from Formula~\eqref{eq:parameters}. As such we omit them from further study.

\subsection{In-degree fit with exogenous communities}

Figures~\ref{fig:dblp-density} and~\ref{fig:cora-density} present the observed and modelled complementary cumulative distribution functions in the communities of the real networks. We estimated all values from our predicted degree sequence for $\rho_i$ values obtained using estimator~\eqref{eq:parameters}. First, looking at Figure~\ref{fig:cora-density}, we observe that all our modelled densities fit well to the observed degree sequences. Our model fits especially well when describing both low- and medium-degree nodes and only overestimated largest in-degree in biggest communities. The other $\rho_i$ estimators often fitted better near distribution means but diverged from real distributions at tails.

Similarly, Figure~\ref{fig:dblp-density} shows that the our analytical prediction matches the empirical densities closely across most clusters, with particularly good agreement at low and medium in-degrees and the highest degrees kept bounded from above. The agreement is strong for all but the two largest communities. We view the remaining deviations as expected: real citation networks exhibit phenomena (ageing, fitness heterogeneity, finite-size effects in the extreme tail) that lie outside the scope of a minimal analytical framework. It should also be emphasised that all curves in Figure~\ref{fig:dblp-density} are produced by a single joint analytical fitting rather than eight independent per-community fits, with parameters coupled through the equations of Section~\ref{sec:experiments}. Consequently, the fit to any one community cannot be improved in isolation without risking deterioration elsewhere, and the quality of the model should be assessed jointly across all communities.

As for fitting to the whole network, we observe that the distribution resulting from the standard 3DSI model is almost identical to the distribution from our model aside from the tail, which suggests that we can use the standard model to approximate the global degree sequence. The parameters for the standard model have been estimated as weighted averages over the communities with probabilities $p_i$ as weights.
Overall, the results indicate that our model fits well to real network data, with the Gini index-based $\rho_i$ estimation producing the high quality results with good tail estimates. The biggest discrepancy happens in high-degree nodes, which our model consistently overestimates. This effect may have many explanations, but in the context of citation networks the occurrence of ageing or attention decay seems the most likely. As our model does not take the attention decay into account, the overestimation is expected for large communities; the problem does not occur for the remaining clusters.

Overall, our model offers a unique possibility to analytically analyse the in-degree sequences of any network with the in-degrees of each community considered separately.

Table~\ref{tab:estimated-parameters} gives the estimated values of the parameters of our model for both networks. We observe that there is a large variance of $\hat{\rho}$ in a single network which confirms that different disciplines tend to cite papers with different ratios of accidentality-to-preferentiality. Our results regarding the Cora network confirm the observations made in~\cite{MROWINSKI2022101341}, where the authors showed that theoretical fields tend to be governed in larger part by accidental citations, while the applied fields have higher preferentiality. This effect is not clearly visible in our results on the DBLP authors dataset, which is likely caused by either different type of topology or large inequality in cluster sizes.
\begin{table}
    \centering
    \caption{Estimated values of the model parameters for all communities in both networks. $\mathcal{G}_i$ denotes the empirical Gini coefficient of the in-degree sequence (used as calibration input for $\hat{\rho}_i$) and $\mathcal{G}_i^{\mathrm{th}}$ denotes the Gini coefficient of the theoretical in-degree sequence produced by the fitted model. Parameters $\hat{\rho_i}$ and $\hat{m_i}$ for the full network have been estimated as weighted averages of the community parameters and used as input parameters when fitting the standard 3DSI model.}
    \small
    \setlength{\tabcolsep}{3pt}
    \begin{tabular}{l|S[table-format=1.3]S[table-format=1.3]S[table-format=1.3]S[table-format=1.3]S[table-format=1.3]||l|S[table-format=3.3]S[table-format=1.3]S[table-format=1.3]S[table-format=1.3]S[table-format=1.3]}
    \hline
        \multicolumn{6}{c}{Cora} & \multicolumn{6}{c}{DBLP authors} \\
        \hline
         Cluster & $\hat{m_i}$ & $\hat{\rho_i}$ & $\hat{p_i}$ & $\mathcal{G}_i$ & $\mathcal{G}_i^{\mathrm{th}}$ & Cluster & $\hat{m_i}$ & $\hat{\rho_i}$ & $\hat{p_i}$  & $\mathcal{G}_i$ & $\mathcal{G}_i^{\mathrm{th}}$\\ \hline
         Case Based       & 1.973 & 0.457 & 0.110 & 0.674 & 0.683 & Comp. network   & 68.567 & 0.639 & 0.030 & 0.758 & 0.706 \\
         Genetic Alg.     & 2.240 & 0.680 & 0.154 & 0.757 & 0.747 & Comp. science   & 142.388 & 0.744 & 0.711 & 0.795 & 0.815 \\
         Neural Nets      & 1.798 & 0.616 & 0.302 & 0.724 & 0.707 & Comp. vision    & 102.756 & 0.818 & 0.050 & 0.797 & 0.787 \\
         Prob. Methods    & 1.946 & 0.534 & 0.157 & 0.679 & 0.700 & Control theory  & 45.579 & 0.746 & 0.075 & 0.820 & 0.677\\
         Reinf. Learning  & 2.338 & 0.710 & 0.080 & 0.753 & 0.758 & Electronic eng. & 34.691 & 0.645 & 0.034 & 0.736 & 0.632\\
         Rule Learning    & 1.955 & 0.528 & 0.066 & 0.711 & 0.700 & Math. opt.      & 67.361 & 0.793 & 0.029 & 0.765 & 0.731\\
         Theory           & 2.166 & 0.677 & 0.130 & 0.728 & 0.742 & Mathematics     & 44.613 & 0.802 & 0.046 & 0.777 & 0.682\\
                          &       &       &       &       &       & Pattern recog.  & 117.940 & 0.528 & 0.025 & 0.801 & 0.750 \\
         \hline
        Full network & 2.005 & 0.605 & 1.000 & 0.721 & {--} &
        Full network & 120.03 & 0.740 & 1.000 & 0.803 & {--} \\
         \hline
    \end{tabular}
    \label{tab:estimated-parameters}
\end{table}
\subsection{In-degree fit with communities identified by the Leiden algorithm}
\begin{figure}
    \centering
    \includegraphics[width=1\linewidth]{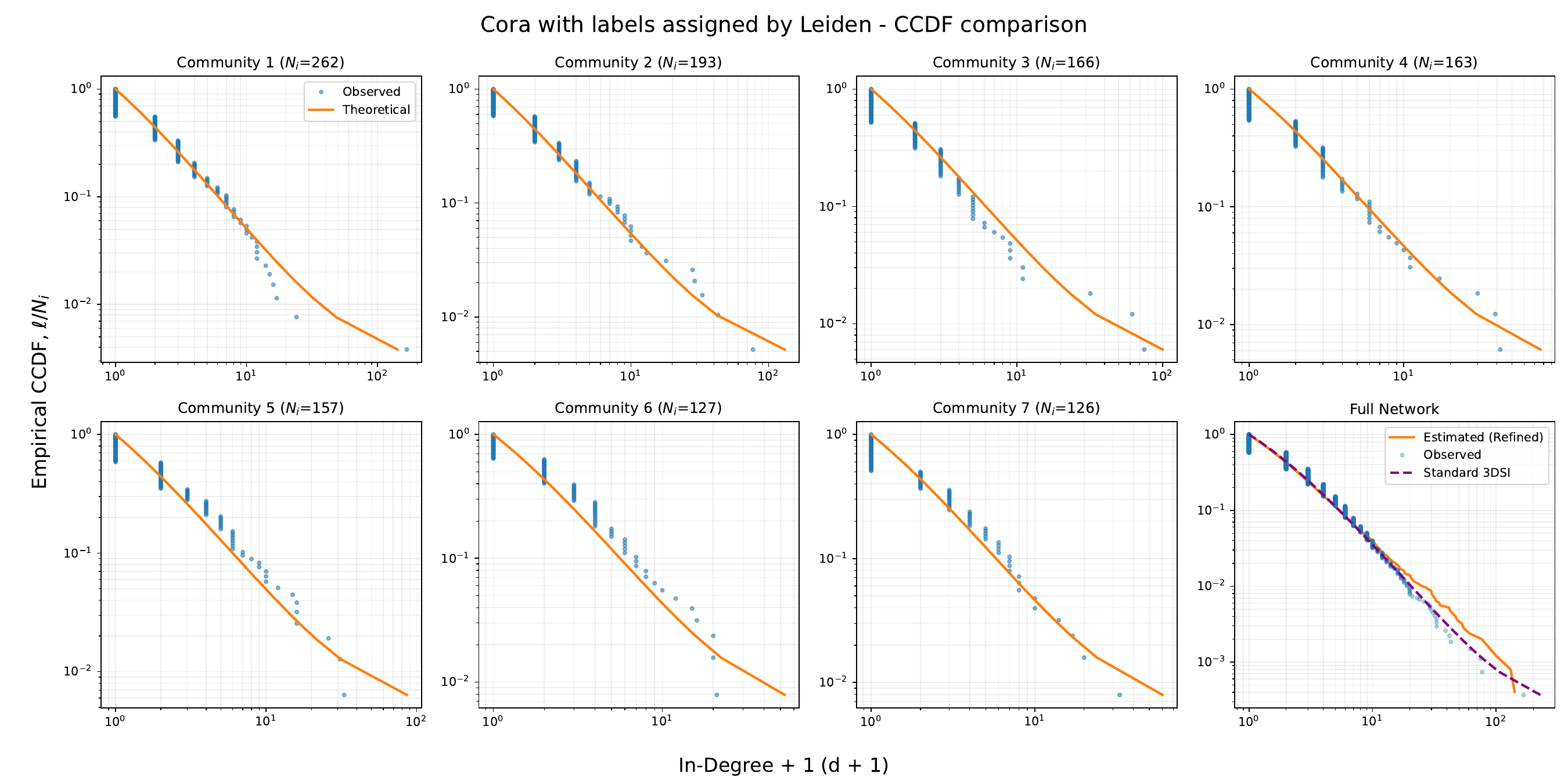}
    \caption{Complementary cumulative distribution functions of the observed and modelled in-degree sequences in the seven largest clusters of the Cora dataset with Leiden-assigned labels. Our model was fitted using the Gini index-based $\rho_i$ estimators. The last plot shows the full in-degree sequences.}
    \label{fig:leiden}
\end{figure}

To confirm that our model works not only with exogenous labels, we repeat the comparison on the Cora dataset but instead of using the ground-truth label vector, we first run the Leiden~\cite{traag-leiden} algorithm to find the  communities. The algorithm is run from a random initial membership until no improvement in modularity is detected. The CCDF comparison on the seven largest found communities, as well as the full network, can be seen in Figure~\ref{fig:leiden}. We selected seven largest communities because the Leiden algorithm found many (100+) clusters, for many of which fitting our model would be pointless due to low sample size and high noise.

Similarly to what we have seen on real data, our analytical predictions match the observed sequences closely for almost all large communities, both around the mode and in the tails. The ageing bias is not visible, likely due to the fairly small sizes of all communities. We find no indication that our model should be limited to exogenous communities; it appears to perform comparably on algorithmically-detected associative communities. There is one caveat -- for extremely small communities, the model cannot be reasonably fitted, and the in-degrees corresponding to such nodes were omitted from the full-network CCDF plot.

\section{Conclusion}\label{sec:conclusion}
We have devised an analytical in-degree sequence model for citation networks that is a direct generalisation of the 3DSI model accounting for the community structure in the graph. We note that it exactly mimics the original 3DSI model with no communities ($k=1$) assuming no loops are allowed. Overall, our model is analytically tractable, yielding closed-form expressions for the in-degree sequence, its Gini index, and the limiting Pareto type II distribution, as well as fits well to real networks. Since it outputs a degree sequence for each community separately, it offers unique insight not seen previously in citation network research.

\paragraph{Limitations.}
As our results indicate, the model fits best to communities that are too young to develop a noticeable ageing or attention decay effect. This directly follows from the fact that our model, developed as the minimal analytical extension of the 3DSI model, has no attention decay parameter and thus is not well suited to modelling networks where the attention decay is prominent. Also, since our derivation, as is standard in the field, allows multi-edges to simplify the derivation, the lack of ageing term may cause the predicted degree of first nodes to exceed the size of the network, especially in high-preferentiality clusters. This effect stems from preferential attachment term predicting more than one edge between some pairs of nodes and is only limited to the few largest in-degree nodes. In particular, our analysis shows that all in-degrees grow asymptotically slower than the network itself.

Another limitation is that our model inherently relies on the community assignments that promote intra-cluster connection but discourage inter-cluster ones. This is expected in citation networks where communities represent scientific domains, but requires careful consideration in other types of networks. This requirement also disqualifies our model from modelling in-degrees in, e.g., dissociative networks.

Notably, our proposition allows only uniformly distributed citations between communities. While intuitively this assumption may be too strong, as preferential attachment likely happens also between communities, we rely on the accidental terms to capture these relationships in order to reduce the number of parameters of the model. It is possible to derive a similar proposition with a separate preferentiality parameter for each pair of communities, but we have decided otherwise to keep the number of parameters linear with respect to $k$. Relaxing this assumption to allow for preferential attachment between arbitrary pairs of communities would require estimating a $k \times k$ matrix of preferentiality parameters, akin to the SBM model. This would lead to a quadratic explosion in the number of parameters, making the model prone to overfitting and difficult to interpret.

\paragraph{Future work.}
We will focus our future works on devising an algorithmic framework for generating benchmark graphs that can be used for community detection, as we believe that our model will be complementary to approaches like SBM or LFR in that it allows for flexible modification of the in-degree sequences inside the communities, allowing for modelling broad families of networks. The degree distribution is only one of many properties of a network; a natural next step is to calibrate and validate the framework against other structural features such as clustering coefficients, assortativity or path-length statistics. It would also be beneficial to consider further generalisations of our model in which accidental edges are also distributed non-homogeneously, or in which the parameter values depend on the size of the network as recent research indicates that interdisciplinary citations trends change non-randomly in time~\cite{rong2025}. Moreover, it would also be interesting to introduce a principled treatment of ageing or attention decay calibrated on time-resolved citation data rather than on static snapshots.

Overall, we believe that our current model proposition is both useful for modelling citation networks as well as offers a new baseline from which to generate even more refined approaches.

\subsection*{Conflict of interest}
The authors certify that they have no affiliations with or involvement in any
organisation or entity with any financial interest or non-financial interest
in the subject matter or materials discussed in this manuscript.

\printcredits  %

\bigskip
\noindent{\small
Please cite this paper as:
Brzozowski, Ł., Gagolewski, M., Siudem, G., Żogała-Siudem, B., The Price–Pareto growth model of networks with community structure, \textit{Journal of Informetrics} \textbf{20}(3), 101850, 2026, \href{https://doi.org/DOI:10.1016/j.joi.2026.101850}{DOI:10.1016/j.joi.2026.101850}.
}


\begin{thebibliography}{42}
\expandafter\ifx\csname natexlab\endcsname\relax\def\natexlab#1{#1}\fi
\providecommand{\url}[1]{\texttt{#1}}
\providecommand{\href}[2]{#2}
\providecommand{\path}[1]{#1}
\providecommand{\DOIprefix}{doi:}
\providecommand{\ArXivprefix}{arXiv:}
\providecommand{\URLprefix}{URL: }
\providecommand{\Pubmedprefix}{pmid:}
\providecommand{\doi}[1]{\href{http://dx.doi.org/#1}{\path{#1}}}
\providecommand{\Pubmed}[1]{\href{pmid:#1}{\path{#1}}}
\providecommand{\bibinfo}[2]{#2}
\ifx\xfnm\relax \def\xfnm[#1]{\unskip,\space#1}\fi
\bibitem[{Albert and Barab\'asi(2002)}]{BA-model}
\bibinfo{author}{Albert, R.}, \bibinfo{author}{Barab\'asi, A.L.},
  \bibinfo{year}{2002}.
\newblock \bibinfo{title}{Statistical mechanics of complex networks}.
\newblock \bibinfo{journal}{Reviews of Modern Physics} \bibinfo{volume}{74},
  \bibinfo{pages}{47--97}.
\newblock \DOIprefix\doi{10.1103/RevModPhys.74.47}.
\bibitem[{Arnold(2015)}]{arnold2015pareto}
\bibinfo{author}{Arnold, B.C.}, \bibinfo{year}{2015}.
\newblock \bibinfo{title}{Pareto Distributions}.
\newblock \bibinfo{edition}{2nd} ed., \bibinfo{publisher}{Chapman and
  Hall/CRC}.
\newblock \DOIprefix\doi{10.1201/b18141}.
\bibitem[{Bertoli-Barsotti et~al.(2024)Bertoli-Barsotti, Gagolewski, Siudem and
  Żogała{-}Siudem}]{OurGiniLorenz}
\bibinfo{author}{Bertoli-Barsotti, L.}, \bibinfo{author}{Gagolewski, M.},
  \bibinfo{author}{Siudem, G.}, \bibinfo{author}{Żogała{-}Siudem, B.},
  \bibinfo{year}{2024}.
\newblock \bibinfo{title}{{G}ini-stable {L}orenz curves and their relation to
  the generalised {P}areto distribution}.
\newblock \bibinfo{journal}{Journal of Informetrics} \bibinfo{volume}{18},
  \bibinfo{pages}{101499}.
\newblock \DOIprefix\doi{10.1016/j.joi.2024.101499}.
\bibitem[{Bianconi and Barabási(2001)}]{BB-model}
\bibinfo{author}{Bianconi, G.}, \bibinfo{author}{Barabási, A.L.},
  \bibinfo{year}{2001}.
\newblock \bibinfo{title}{Competition and multiscaling in evolving networks}.
\newblock \bibinfo{journal}{Europhysics Letters} \bibinfo{volume}{54},
  \bibinfo{pages}{436}.
\newblock \DOIprefix\doi{10.1209/epl/i2001-00260-6}.
\bibitem[{Blei et~al.(2003)Blei, Ng and Jordan}]{lda-model}
\bibinfo{author}{Blei, D.M.}, \bibinfo{author}{Ng, A.Y.},
  \bibinfo{author}{Jordan, M.I.}, \bibinfo{year}{2003}.
\newblock \bibinfo{title}{Latent {D}irichlet allocation}.
\newblock \bibinfo{journal}{Journal of Machine Learning Research}
  \bibinfo{volume}{3}, \bibinfo{pages}{993–1022}.
\bibitem[{Blondel et~al.(2008)Blondel, Guillaume, Lambiotte and
  Lefebvre}]{blondel-louvain}
\bibinfo{author}{Blondel, V.D.}, \bibinfo{author}{Guillaume, J.L.},
  \bibinfo{author}{Lambiotte, R.}, \bibinfo{author}{Lefebvre, E.},
  \bibinfo{year}{2008}.
\newblock \bibinfo{title}{Fast unfolding of communities in large networks}.
\newblock \bibinfo{journal}{Journal of Statistical Mechanics: Theory and
  Experiment} \bibinfo{volume}{2008}, \bibinfo{pages}{P10008}.
\newblock \DOIprefix\doi{10.1088/1742-5468/2008/10/P10008}.
\bibitem[{Cascarina(2023)}]{Cascarina2023}
\bibinfo{author}{Cascarina, S.M.}, \bibinfo{year}{2023}.
\newblock \bibinfo{title}{Self-referencing rates in biological disciplines}.
\newblock \bibinfo{journal}{Frontiers in Research Metrics and Analytics}
  \bibinfo{volume}{8}, \bibinfo{pages}{1215401}.
\newblock \DOIprefix\doi{10.3389/frma.2023.1215401}.
\bibitem[{Chang and Blei(2009)}]{rtm-model}
\bibinfo{author}{Chang, J.}, \bibinfo{author}{Blei, D.}, \bibinfo{year}{2009}.
\newblock \bibinfo{title}{Relational topic models for document networks}, in:
  \bibinfo{editor}{van Dyk, D.}, \bibinfo{editor}{Welling, M.} (Eds.),
  \bibinfo{booktitle}{Proceedings of the Twelfth International Conference on
  Artificial Intelligence and Statistics}, pp. \bibinfo{pages}{81--88}.
\bibitem[{Chen et~al.(2014)Chen, Chuang and Chiu}]{chen2014}
\bibinfo{author}{Chen, Y.L.}, \bibinfo{author}{Chuang, C.H.},
  \bibinfo{author}{Chiu, Y.T.}, \bibinfo{year}{2014}.
\newblock \bibinfo{title}{Community detection based on social interactions in a
  social network}.
\newblock \bibinfo{journal}{Journal of the Association for Information Science
  and Technology} \bibinfo{volume}{65}, \bibinfo{pages}{539--550}.
\newblock \DOIprefix\doi{10.1002/asi.22986}.
\bibitem[{Dai et~al.(2021)Dai, Chen, Wan, Liu, Gong and Wang}]{Dai2021}
\bibinfo{author}{Dai, C.}, \bibinfo{author}{Chen, Q.}, \bibinfo{author}{Wan,
  T.}, \bibinfo{author}{Liu, F.}, \bibinfo{author}{Gong, Y.},
  \bibinfo{author}{Wang, Q.}, \bibinfo{year}{2021}.
\newblock \bibinfo{title}{Literary runaway: {I}ncreasingly more references
  cited per academic research article from 1980 to 2019}.
\newblock \bibinfo{journal}{PLOS ONE} \bibinfo{volume}{16},
  \bibinfo{pages}{e0255849}.
\newblock \DOIprefix\doi{10.1371/journal.pone.0255849}.
\bibitem[{Decelle et~al.(2011)Decelle, Krzakala, Moore and
  Zdeborov{\'a}}]{decelle2011inference}
\bibinfo{author}{Decelle, A.}, \bibinfo{author}{Krzakala, F.},
  \bibinfo{author}{Moore, C.}, \bibinfo{author}{Zdeborov{\'a}, L.},
  \bibinfo{year}{2011}.
\newblock \bibinfo{title}{Inference and phase transitions in the detection of
  modules in sparse networks}.
\newblock \bibinfo{journal}{Physical Review Letters} \bibinfo{volume}{107},
  \bibinfo{pages}{065701}.
\bibitem[{Dorfman(1979)}]{gini-formulas}
\bibinfo{author}{Dorfman, R.}, \bibinfo{year}{1979}.
\newblock \bibinfo{title}{A formula for the {G}ini coefficient}.
\newblock \bibinfo{journal}{The Review of Economics and Statistics}
  \bibinfo{volume}{61}, \bibinfo{pages}{146--149}.
\bibitem[{Drobyshevskiy and Turdakov(2019)}]{Drobyshevskiy-survey}
\bibinfo{author}{Drobyshevskiy, M.}, \bibinfo{author}{Turdakov, D.},
  \bibinfo{year}{2019}.
\newblock \bibinfo{title}{Random graph modeling: {A} survey of the concepts}.
\newblock \bibinfo{journal}{ACM Computing Surveys} \bibinfo{volume}{52},
  \bibinfo{pages}{131}.
\newblock \DOIprefix\doi{10.1145/3369782}.
\bibitem[{Fortunato(2010)}]{fortunato-survey}
\bibinfo{author}{Fortunato, S.}, \bibinfo{year}{2010}.
\newblock \bibinfo{title}{Community detection in graphs}.
\newblock \bibinfo{journal}{Physics Reports} \bibinfo{volume}{486},
  \bibinfo{pages}{75--174}.
\newblock \DOIprefix\doi{10.1016/j.physrep.2009.11.002}.
\bibitem[{Fronczak et~al.(2013)Fronczak, Fronczak and Bujok}]{fronczak-sbm}
\bibinfo{author}{Fronczak, P.}, \bibinfo{author}{Fronczak, A.},
  \bibinfo{author}{Bujok, M.}, \bibinfo{year}{2013}.
\newblock \bibinfo{title}{Exponential random graph models for networks with
  community structure}.
\newblock \bibinfo{journal}{Physical Review E} \bibinfo{volume}{88},
  \bibinfo{pages}{032810}.
\newblock \DOIprefix\doi{10.1103/PhysRevE.88.032810}.
\bibitem[{Gautschi(1959)}]{Gautschi1959}
\bibinfo{author}{Gautschi, W.}, \bibinfo{year}{1959}.
\newblock \bibinfo{title}{Some elementary inequalities relating to the {Gamma}
  and incomplete {Gamma} function}.
\newblock \bibinfo{journal}{Journal of Mathematics and Physics}
  \bibinfo{volume}{38}, \bibinfo{pages}{77--81}.
\newblock \DOIprefix\doi{10.1002/sapm195938177}.
\bibitem[{Golosovsky(2017)}]{golosovsky2017}
\bibinfo{author}{Golosovsky, M.}, \bibinfo{year}{2017}.
\newblock \bibinfo{title}{Growing complex network of citations of scientific
  papers: {M}odeling and measurements}.
\newblock \bibinfo{journal}{Physical Review E} \bibinfo{volume}{96},
  \bibinfo{pages}{032306}.
\bibitem[{Gradshteyn and Ryzhik(2007)}]{gradshtein-definite-integrals}
\bibinfo{author}{Gradshteyn, I.S.}, \bibinfo{author}{Ryzhik, I.M.},
  \bibinfo{year}{2007}.
\newblock \bibinfo{title}{3–4 --- {D}efinite integrals of elementary
  functions}, in: \bibinfo{editor}{Jeffrey, A.}, \bibinfo{editor}{Zwillinger,
  D.}, \bibinfo{editor}{Gradshteyn, I.}, \bibinfo{editor}{Ryzhik, I.} (Eds.),
  \bibinfo{booktitle}{Table of Integrals, Series, and Products}.
  \bibinfo{edition}{7th} ed.. \bibinfo{publisher}{Academic Press},
  \bibinfo{address}{Boston}, pp. \bibinfo{pages}{247--617}.
\newblock \DOIprefix\doi{10.1016/B978-0-08-047111-2.50013-3}.
\bibitem[{Harzing(2010)}]{Harzing2010}
\bibinfo{author}{Harzing, A.}, \bibinfo{year}{2010}.
\newblock \bibinfo{title}{Citation analysis across disciplines: {T}he impact of
  different data sources and citation metrics}.
\newblock \bibinfo{howpublished}{White paper, Anne‐Will Harzing's site}.
\newblock \URLprefix
  \url{https://harzing.com/publications/white-papers/citation-analysis-across-disciplines}.
  \bibinfo{note}{available online, accessed 30/09/2025}.
\bibitem[{Holland et~al.(1983)Holland, Laskey and Leinhardt}]{holland-sbm}
\bibinfo{author}{Holland, P.W.}, \bibinfo{author}{Laskey, K.B.},
  \bibinfo{author}{Leinhardt, S.}, \bibinfo{year}{1983}.
\newblock \bibinfo{title}{Stochastic blockmodels: First steps}.
\newblock \bibinfo{journal}{Social Networks} \bibinfo{volume}{5},
  \bibinfo{pages}{109--137}.
\newblock \DOIprefix\doi{10.1016/0378-8733(83)90021-7}.
\bibitem[{Kipf and Welling(2017)}]{kipfgcn}
\bibinfo{author}{Kipf, T.N.}, \bibinfo{author}{Welling, M.},
  \bibinfo{year}{2017}.
\newblock \bibinfo{title}{Semi-supervised classification with graph
  convolutional networks}, in: \bibinfo{booktitle}{5th International Conference
  on Learning Representations, {ICLR} 2017}.
\bibitem[{Lancichinetti et~al.(2008)Lancichinetti, Fortunato and
  Radicchi}]{lancichinetti-lfr}
\bibinfo{author}{Lancichinetti, A.}, \bibinfo{author}{Fortunato, S.},
  \bibinfo{author}{Radicchi, F.}, \bibinfo{year}{2008}.
\newblock \bibinfo{title}{Benchmark graphs for testing community detection
  algorithms}.
\newblock \bibinfo{journal}{Physical Review E} \bibinfo{volume}{78}.
\newblock \DOIprefix\doi{10.1103/physreve.78.046110}.
\bibitem[{Leskovec and Krevl(2014)}]{snapnets}
\bibinfo{author}{Leskovec, J.}, \bibinfo{author}{Krevl, A.},
  \bibinfo{year}{2014}.
\newblock \bibinfo{title}{{SNAP Datasets}: {Stanford} large network dataset
  collection}.
\newblock \bibinfo{howpublished}{\url{http://snap.stanford.edu/data}}.
\bibitem[{Li et~al.(2020)Li, Yu, Li, Zhang, Zhao, Rong, Cheng and Huang}]{gvae}
\bibinfo{author}{Li, J.}, \bibinfo{author}{Yu, J.}, \bibinfo{author}{Li, J.},
  \bibinfo{author}{Zhang, H.}, \bibinfo{author}{Zhao, K.},
  \bibinfo{author}{Rong, Y.}, \bibinfo{author}{Cheng, H.},
  \bibinfo{author}{Huang, J.}, \bibinfo{year}{2020}.
\newblock \bibinfo{title}{{D}irichlet graph variational autoencoder}, in:
  \bibinfo{editor}{Larochelle, H.}, \bibinfo{editor}{Ranzato, M.},
  \bibinfo{editor}{Hadsell, R.}, \bibinfo{editor}{Balcan, M.},
  \bibinfo{editor}{Lin, H.} (Eds.), \bibinfo{booktitle}{Advances in Neural
  Information Processing Systems}, pp. \bibinfo{pages}{5274--5283}.
\bibitem[{Medo et~al.(2011)Medo, Cimini and Gualdi}]{medo2011}
\bibinfo{author}{Medo, M.}, \bibinfo{author}{Cimini, G.},
  \bibinfo{author}{Gualdi, S.}, \bibinfo{year}{2011}.
\newblock \bibinfo{title}{Temporal dynamics of popularity: {T}he case of
  scientific papers}.
\newblock \bibinfo{journal}{Physical Review Letters} \bibinfo{volume}{107},
  \bibinfo{pages}{238701}.
\bibitem[{Milojevi{\'{c}}(2025)}]{Milojevic2025}
\bibinfo{author}{Milojevi{\'{c}}, S.}, \bibinfo{year}{2025}.
\newblock \bibinfo{title}{Science of science}.
\newblock \bibinfo{journal}{Scientometrics} \bibinfo{volume}{130},
  \bibinfo{pages}{3195--3211}.
\newblock \DOIprefix\doi{10.1007/s11192-025-05322-1}.
\bibitem[{Mrowinski et~al.(2022)Mrowinski, Gagolewski and
  Siudem}]{MROWINSKI2022101341}
\bibinfo{author}{Mrowinski, M.J.}, \bibinfo{author}{Gagolewski, M.},
  \bibinfo{author}{Siudem, G.}, \bibinfo{year}{2022}.
\newblock \bibinfo{title}{Accidentality in journal citation patterns}.
\newblock \bibinfo{journal}{Journal of Informetrics} \bibinfo{volume}{16},
  \bibinfo{pages}{101341}.
\newblock \DOIprefix\doi{10.1016/j.joi.2022.101341}.
\bibitem[{Newman(2018)}]{newman-networks}
\bibinfo{author}{Newman, M.}, \bibinfo{year}{2018}.
\newblock \bibinfo{title}{Networks}.
\newblock \bibinfo{publisher}{Oxford University Press}.
\newblock \DOIprefix\doi{10.1093/oso/9780198805090.001.0001}.
\bibitem[{Olejniczak et~al.(2022)Olejniczak, Savage and
  Wheeler}]{OlejniczakSavageWheeler2022}
\bibinfo{author}{Olejniczak, A.J.}, \bibinfo{author}{Savage, W.E.},
  \bibinfo{author}{Wheeler, R.}, \bibinfo{year}{2022}.
\newblock \bibinfo{title}{The rhythms of scholarly publication: {S}uggestions
  to enhance bibliometric comparisons across disciplines}.
\newblock \bibinfo{journal}{Frontiers in Research Metrics and Analytics}
  \bibinfo{volume}{7}.
\newblock \DOIprefix\doi{10.3389/frma.2022.812312}.
\bibitem[{Parolo et~al.(2015)Parolo, Pan, Ghosh, Huberman, Kaski and
  Fortunato}]{parolo2015}
\bibinfo{author}{Parolo, P.D.B.}, \bibinfo{author}{Pan, R.K.},
  \bibinfo{author}{Ghosh, R.}, \bibinfo{author}{Huberman, B.A.},
  \bibinfo{author}{Kaski, K.}, \bibinfo{author}{Fortunato, S.},
  \bibinfo{year}{2015}.
\newblock \bibinfo{title}{Attention decay in science}.
\newblock \bibinfo{journal}{Journal of Informetrics} \bibinfo{volume}{9},
  \bibinfo{pages}{734--745}.
\bibitem[{Price(1965)}]{deSollaPrice1965}
\bibinfo{author}{Price, D.}, \bibinfo{year}{1965}.
\newblock \bibinfo{title}{Networks of scientific papers}.
\newblock \bibinfo{journal}{Science} \bibinfo{volume}{149},
  \bibinfo{pages}{510--515}.
\newblock \DOIprefix\doi{10.1126/science.149.3683.510}.
\bibitem[{Rong et~al.(2025)Rong, Chen, Ma and Koch}]{rong2025}
\bibinfo{author}{Rong, G.}, \bibinfo{author}{Chen, Y.}, \bibinfo{author}{Ma,
  F.}, \bibinfo{author}{Koch, T.}, \bibinfo{year}{2025}.
\newblock \bibinfo{title}{Exploring interdisciplinary research trends through
  critical years for interdisciplinary citation}.
\newblock \bibinfo{journal}{Journal of Informetrics} \bibinfo{volume}{19},
  \bibinfo{pages}{101726}.
\newblock \DOIprefix\doi{10.1016/j.joi.2025.101726}.
\bibitem[{Scarselli et~al.(2009)Scarselli, Gori, Tsoi, Hagenbuchner and
  Monfardini}]{scarselli2009}
\bibinfo{author}{Scarselli, F.}, \bibinfo{author}{Gori, M.},
  \bibinfo{author}{Tsoi, A.C.}, \bibinfo{author}{Hagenbuchner, M.},
  \bibinfo{author}{Monfardini, G.}, \bibinfo{year}{2009}.
\newblock \bibinfo{title}{The graph neural network model}.
\newblock \bibinfo{journal}{IEEE Transactions on Neural Networks}
  \bibinfo{volume}{20}, \bibinfo{pages}{61--80}.
\newblock \DOIprefix\doi{10.1109/TNN.2008.2005605}.
\bibitem[{Sen et~al.(2008)Sen, Namata, Bilgic, Getoor, Gallagher and
  Eliassi-Rad}]{cora-dataset}
\bibinfo{author}{Sen, P.}, \bibinfo{author}{Namata, G.M.},
  \bibinfo{author}{Bilgic, M.}, \bibinfo{author}{Getoor, L.},
  \bibinfo{author}{Gallagher, B.}, \bibinfo{author}{Eliassi-Rad, T.},
  \bibinfo{year}{2008}.
\newblock \bibinfo{title}{Collective classification in network data}.
\newblock \bibinfo{journal}{AI Magazine} \bibinfo{volume}{29},
  \bibinfo{pages}{93--106}.
\bibitem[{Simkin and Roychowdhury(2007)}]{simkin}
\bibinfo{author}{Simkin, M.}, \bibinfo{author}{Roychowdhury, V.},
  \bibinfo{year}{2007}.
\newblock \bibinfo{title}{A mathematical theory of citing}.
\newblock \bibinfo{journal}{Journal of the Association for Information Science
  and Technology} \bibinfo{volume}{58}, \bibinfo{pages}{1661--1673}.
\newblock \DOIprefix\doi{10.1002/asi.20653}.
\bibitem[{Siudem et~al.(2022)Siudem, Nowak and Gagolewski}]{pricepareto2}
\bibinfo{author}{Siudem, G.}, \bibinfo{author}{Nowak, P.},
  \bibinfo{author}{Gagolewski, M.}, \bibinfo{year}{2022}.
\newblock \bibinfo{title}{Power laws, the {P}rice model, and the {P}areto
  type-2 distribution}.
\newblock \bibinfo{journal}{Physica A: Statistical Mechanics and its
  Applications} \bibinfo{volume}{606}, \bibinfo{pages}{128059}.
\newblock \DOIprefix\doi{10.1016/j.physa.2022.128059}.
\bibitem[{Siudem et~al.(2020)Siudem, Żogała{-}Siudem, Cena and
  Gagolewski}]{3dsi-pnas}
\bibinfo{author}{Siudem, G.}, \bibinfo{author}{Żogała{-}Siudem, B.},
  \bibinfo{author}{Cena, A.}, \bibinfo{author}{Gagolewski, M.},
  \bibinfo{year}{2020}.
\newblock \bibinfo{title}{Three dimensions of scientific impact}.
\newblock \bibinfo{journal}{Proceedings of the National Academy of Sciences of
  the United States of America (PNAS)} \bibinfo{volume}{117},
  \bibinfo{pages}{13896--13900}.
\newblock \DOIprefix\doi{10.1073/pnas.2001064117}.
\bibitem[{Tang et~al.(2008)Tang, Zhang, Yao, Li, Zhang and Su}]{dblp-dataset}
\bibinfo{author}{Tang, J.}, \bibinfo{author}{Zhang, J.}, \bibinfo{author}{Yao,
  L.}, \bibinfo{author}{Li, J.}, \bibinfo{author}{Zhang, L.},
  \bibinfo{author}{Su, Z.}, \bibinfo{year}{2008}.
\newblock \bibinfo{title}{Arnetminer: Extraction and mining of academic social
  networks}, in: \bibinfo{booktitle}{KDD'08}, pp. \bibinfo{pages}{990--998}.
\bibitem[{Tian et~al.(2023)Tian, Li and Mao}]{tian2023}
\bibinfo{author}{Tian, Y.}, \bibinfo{author}{Li, G.}, \bibinfo{author}{Mao,
  J.}, \bibinfo{year}{2023}.
\newblock \bibinfo{title}{Predicting the evolution of scientific communities by
  interpretable machine learning approaches}.
\newblock \bibinfo{journal}{Journal of Informetrics} \bibinfo{volume}{17},
  \bibinfo{pages}{101399}.
\newblock \DOIprefix\doi{10.1016/j.joi.2023.101399}.
\bibitem[{Traag et~al.(2019)Traag, Waltman and van Eck}]{traag-leiden}
\bibinfo{author}{Traag, V.}, \bibinfo{author}{Waltman, L.},
  \bibinfo{author}{van Eck, N.J.}, \bibinfo{year}{2019}.
\newblock \bibinfo{title}{From {L}ouvain to {L}eiden: {G}uaranteeing
  well-connected communities}.
\newblock \bibinfo{journal}{Scientific Reports} \bibinfo{volume}{9},
  \bibinfo{pages}{5233}.
\newblock \DOIprefix\doi{10.1038/s41598-019-41695-z}.
\bibitem[{Wang et~al.(2013)Wang, Song and Barab{\'a}si}]{wang2013}
\bibinfo{author}{Wang, D.}, \bibinfo{author}{Song, C.},
  \bibinfo{author}{Barab{\'a}si, A.L.}, \bibinfo{year}{2013}.
\newblock \bibinfo{title}{Quantifying long-term scientific impact}.
\newblock \bibinfo{journal}{Science} \bibinfo{volume}{342},
  \bibinfo{pages}{127--132}.
\bibitem[{Zhang and Tay(2016)}]{gscaler}
\bibinfo{author}{Zhang, J.W.}, \bibinfo{author}{Tay, Y.C.},
  \bibinfo{year}{2016}.
\newblock \bibinfo{title}{{GSCALER:} {S}ynthetically scaling a given graph},
  in: \bibinfo{editor}{Pitoura, E.}, \bibinfo{editor}{Maabout, S.},
  \bibinfo{editor}{Koutrika, G.}, \bibinfo{editor}{Marian, A.},
  \bibinfo{editor}{Tanca, L.}, \bibinfo{editor}{Manolescu, I.},
  \bibinfo{editor}{Stefanidis, K.} (Eds.), \bibinfo{booktitle}{Proceedings of
  the 19th International Conference on Extending Database Technology, {EDBT}
  2016, Bordeaux, France, March 15-16, 2016, Bordeaux, France, March 15-16,
  2016}, pp. \bibinfo{pages}{53--64}.
\newblock \DOIprefix\doi{10.5441/002/EDBT.2016.08}.

\end{thebibliography}
\end{document}